\newcommand{\diracslash}[1]{#1\llap{/\kern2pt}}

\newcommand{\be}{\begin{equation}}
\newcommand{\ee}{\end{equation}}
\newcommand{\bea}{\begin{eqnarray}}
\newcommand{\eea}{\end{eqnarray}}
\newcommand{\ba}[1]{\begin{array}{#1}}
\newcommand{\ea}{\end{array}}

\newcommand{\bt}{\begin{tabular}}
\newcommand{\et}{\end{tabular}}

\newcommand{\beas}{\begin{eqnarray*}}
\newcommand{\eeas}{\end{eqnarray*}}

\documentclass[preprint,prd,aps,floats,nofootinbib,floatfix]{revtex4}
\usepackage{graphicx}
\begin{document}

\title{Open bottom mesons in asymmetric nuclear matter
in presence of strong magnetic fields}
\author{Nikhil Dhale}
\email{dhalenikhil07@gmail.com}
\affiliation{Department of Physics, Indian Institute of Technology, Delhi,
Hauz Khas, New Delhi -- 110 016, India}

\author{Sushruth Reddy P}
\email{sushruth.p7@gmail.com}
\affiliation{Department of Physics, Indian Institute of Technology, Delhi,
Hauz Khas, New Delhi -- 110 016, India}

\author{Amal Jahan CS}
\email{amaljahan@gmail.com}
\affiliation{Department of Physics, Indian Institute of Technology, Delhi,
Hauz Khas, New Delhi -- 110 016, India}

\author{Amruta Mishra}
\email{amruta@physics.iitd.ac.in}
\affiliation{Department of Physics, Indian Institute of Technology, Delhi,
Hauz Khas, New Delhi -- 110 016, India}

\begin{abstract}
The modifications of the masses of the $B$ and $\bar B$ mesons
in asymmetric nuclear matter in the presence of strong magnetic fields,
are investigated using a chiral effective model.
The medium modifications of these open bottom mesons
arise due to their interactions with the scalar mesons and the
nucleons. In the magnetized nuclear matter,
the proton has contributions from the Landau levels. 
In the chiral effective model, the masses of the $B$ and $\bar B$ 
mesons are calculated from the leading term, namely the vectorial
Weinberg Tomozawa term as well as from the next to leading order 
contributions, i.e., due to the scalar exchange and the range terms.
Due to the Weinberg-Tomozawa term, the $\bar B$ mesons experience 
an attractive interaction in the symmetric nuclear matter, 
whereas the $B$ mesons have a repulsive interaction. 
Inclusion of the contributions from the scalar exchange and the range terms
as well, leads to drop of the masses of both $B$ and $\bar B$ mesons.
The effect of the isospin asymmetry breaks the mass degeneracy
of the $B^+$ and $B^0$ (as well as of the $B^-$ and $\bar {B^0}$)
mesons, and its effect is observed to be large at high densities.
The effects of anomalous magnetic moments of the nucleons
are taken into account in the present study of the masses 
of the open bottom mesons in magnetized nuclear matter.
\end{abstract}
\maketitle

\def\bfm#1{\mbox{\boldmath $#1$}}

\section{Introduction}
The study of properties of hadrons in the hadronic medium 
is an intense field of research in high energy physics.
This is due to the relevance of the subject
in ultra-relativistic heavy-ion collision experiments,
where the experimental observables are affected
by the medium modifications of the hadrons.
At high energy heavy ion collision experiments,
huge magnetic fields
are created ($eB \sim 6 m_\pi^2$ at RHIC, BNL, and
$eB \sim 15 m_\pi^2$ at LHC, CERN)  \cite{kharzeev,skokov}. 
It is thus important to
study the effects of the magnetic fields on the hadrons
in the strongly interacting matter resulting from
the high energy nuclear collisions. The study of
magnetic fields is also important
for astrophysical objects like the neutron
stars with magnetic fields/magnetars
\cite{magnetarbb,somenathprl,broderick1,broderick2,Wei,mao, 
somenathprd,ambhasmag}.
The presence of magnetic fields also lead to 
novel effects like chiral magnetic effect and the
inverse magnetic catalysis \cite{mag_catalysis},
which has made the
study of effects of the magnetic fields in strongly
interacting matter even more interesting.
The magnetic fields created in non-central relativistic 
heavy ion collision experiments drop rapidly after the collision, 
creating induced currents, which tend to slow down the decay of the
magnetic field \cite{tuchin2011,tuchin2010,Ajit2017}.
Also at later times, the matter effects are believed to play
an important role in the time evolution of the magnetic
field, with the effect of slowing down the decay of the magnetic field
\cite{tuchin2013}. The time evolution of the magnetic 
field in the heavy ion collision experiments is still 
an open problem, which needs careful investigation 
of the electrical conductivity of the medium
as well as solutions of the magnetohydrodynamic equations.
The fact that there are huge magnetic fields
created in relativistic heavy ion collision experiments,
has initiated number of studies of the hadrons,
including those of heavy flavour mesons, e.g.,
open charm and open bottom mesons 
\cite{machado_1,machado_2,gubler,dmeson_mag}, 
as well as heavy quarkonium states, e.g, the charmonium states
\cite{charmonium_mag_lee}, in the presence of magnetic fields.

The masses of the heavy (charm and bottom)-light  mesons, 
e.g. the $D$ and $B$ mesons, are modified  appreciably
in the hadronic medium. This is because, due to the 
presence of the light quark (antiquark) (along
with the heavy charm (bottom) antiquark (quark)),
these mesons interact with the light quark condensates
\cite{arata},
which are modified significantly in the medium.
The effects of these medium modifications should
show in observables, e.g. the production and propagation
of these particles. There have been various approaches
to study the in-medium properties of these heavy flavour mesons,
e.g., QCD sum rule approach 
\cite{arata,qcdsum08,epja2011_Hilger,jpg2010_Hilger}
(where the in-medium masses are calculated 
from the medium modifications of the quark and gluon condensates),
the Quark Meson Coupling (QMC) model \cite{qmc_D}, 
in which the quarks interact via scalar and vector mesons
\cite{qmc}, the coupled channel approach 
\cite{ltolos,ljhs,mizutani,HL}, 
using effective hadronic models, e.g. chiral effective model 
\cite{amdmeson, amarindamprc, amarvdmesonTprc, amarvepja},
studies using heavy quark symmetry
and $\bar D$ ($B$) interaction with nucleon using pion exchange
\cite{Yasui_Sudoh_pion}, and, using a heavy meson effective theory
with $O(1/M)$ corrections \cite{Yasui_Sudoh_heavy_meson_Eff_th},
There has also been a study of the heavy flavour meson (heavy quark)
embedded as an impurity in the nuclear matter (quark matter)
by Yasui and Sudoh, which have
very similar behaviour at large mass (of the meson/quark) limit
\cite{Yasui_Sudoh_heavy_particle_impurity}. 
The open charm and open bottom mesons studied
using pion exchange for the heavy meson--nucleon 
interaction \cite{Yasui_Sudoh_pion}, 
give attractive interactions of the
$\bar D$ and $B$ mesons in nuclear matter, which predict possibility
of the bound states of these mesons with the nuclei \cite{qmc_D}.
Due to attractive interaction of $J/\psi$ in nuclear matter
\cite{leeko,krein_jpsi,amarvdmesonTprc,amarvepja},
the bound states of the $J/\psi$ to nuclei have also been
predicted within the QMC model \cite{krein_17}.
The heavy flavour hadrons in nuclear matter 
\cite{Hosaka_Prog_Part_Nucl_Phys} have gained 
considerable interest in the recent years, due to the
possibility of heavy hadrons, e.g., $\bar D$,
$B$, $J/\psi$ bound in nuclei, as well as $\Lambda_c^+$ and $\Lambda_b$
hypernuclei \cite{qmc_lambda_c_b_hypernuclei}.
The open heavy flavour (charm and bottom) scalar as well as 
vector mesons, due to the presence of a light quark (antiquark),
also undergo appreciable mass modifications in the medium,
similar to the pseudoscalar $D(\bar D)$ and $B(\bar B)$ mesons. 
These mesons have been studied using QCD sum rule approach
\cite{Wang_heavy_scalar,Wang_heavy_meson,Hilger_scalar_open_charm,
Hilger_sc_PS_open_charm,arvind_heavy_mesons_QSR}
as well as within the coupled channel framework
\cite{tolos_heavy_mesons}. 

In the present work, the mass modifications of the $B$ 
and $\bar B$ mesons are studied in symmetric as well as asymmetric
nuclear matter in the 
presence of strong magnetic fields, using a chiral effective model.
The model is a generalization of a chiral SU(3) model to include
the interactions of the charm and bottom mesons with the 
light hadronic sector. The chiral SU(3) model, based
on a nonlinear realization of chiral symmetry, 
has been used to study nuclear matter,
finite nuclei \cite{paper3}, hyperonic matter \cite{kristof1}, 
vector mesons \cite{hartree},
kaons and antikaons \cite{kaon_antikaon,isoamss,isoamss1,isoamss2},
as well as to study the charge neutral
matter present in the interior of the (proto) neutron stars
\cite{pneutronstar}.
The chiral SU(3) model, generalized to the charm and bottom sectors,
and has been used to study the $D$ and $\bar D$ mesons
\cite{amarindamprc, amarvdmesonTprc, amarvepja},
the strange charm mesons (the $D_s$ mesons)
\cite{DP_AM_Ds}, the $B$, $\bar B$ mesons \cite{DP_AM_bbar}, and,
the strange bottom mesons, i.e., the $B_s$ mesons \cite{DP_AM_Bs},
as arising from their interactions to the baryons 
and the scalar mesons. Within effective hadronic model, 
the broken scale symmetry
of QCD \cite{paper3, kristof1,sche1} has also been incorporated 
through a scalar dilaton field which mimicks the
gluon condensates of QCD. 
The mass modifications of the heavy quarkonium mesons, 
e.g., charmonium \cite{amarvdmesonTprc,amarvepja} 
and bottomonium states \cite{AM_DP_upsilon}, which do not contain
any light quark/antiquark, arise due to their
interactions with the gluon condensate of QCD.
These are calculated within the chiral effective model from
the medium modifications of the dilaton field in the nuclear matter.
The in-medium partial decay widths 
of the charmonium states ($J/\psi$ and the excited states
of charmonium, e.g., $\psi(3686)$, $\psi(3770)$,
as well as $\chi_c$) to $D\bar D$ have been studied
in the literature, using a light quark pair creation model, 
namely $^3P_0$ model \cite{3p0_1,3p0_2}. These have been
calculated by assuming mass modifications
of the $D$ and $\bar D$ mesons \cite{friman},
but without accounting for any mass modifications 
of the charmonium states. Using the $^3P_0$ model, 
from the medium modifications of the masses of the
open charm mesons and the charmonium states,
calculated within the chiral effective model,
the partial decay widths of the charmonium states
to $D\bar D$ in hadronic matter, 
have also been studied \cite{amarvepja}.
Subsequently, the in-medium charmonium decay widths
have been studied, using a field theoretic model
of composite hadrons \cite{amspmwg}. 
The partial decay widths of the $\Upsilon$-states to $B\bar B$,
in hadronic matter have later been studied using the field theoretic
model of composite hadrons with quark constituents
\cite{amspm_upsilon}, using the in-medium masses of the
$B$ and $\bar B$ \cite{DP_AM_bbar}, as well as the bottomonium
states \cite{AM_DP_upsilon}, as calculated within the 
chiral effective model. In the mean field approximation,
the values of the scalar mesons, $\sigma$, $\zeta$ and $\delta$,
as calculated from their equations of motion,
are related to the light quark condensates, and the
dilaton field is related to the gluon condensate in the medium,
which have been used to study the properties of the light vector 
mesons ($\omega$, $\rho$ and $\phi$) \cite{am_vecmeson},
as well as the charmonium states, $J/\psi$ and $\eta_c$
\cite{amarvjpsi_qsr} using the QCD sum rule approach.
The $D$ mesons in the presence of an external magnetic field
have been studied in the literature,
within the QCD sum rule approach,
accounting for the mixing of the pesudoscalar mesons
and the vector mesons, as well as the
Landau quantization effects for the charged $D$ mesons \cite{gubler}.
Using a semiclassical approach \cite{machado_1}, 
the open charm and open bottom mesons have been
studied in the presence of an external magnetic field,
where the mass of the open charm (bottom)
meson is due to the interaction of the magnetic field
to the spin of the quarks. 
Different spin orientations of the heavy-light quark-antiquark 
systems were observed to have different masses in the
presence of the magnetic field.
The $D$ and $B$ mesons were observed to have lowering of their masses
(arising from the mass reductions of specific spin orientations)
\cite{machado_1}. 
Further, M1 transitions become dominant in the presence 
of strong magnetic fields, leading to the mixing 
of the spin--0 states $D$ ($B$) to the spin--1 states $D^*$ ($B^*$).
The masses of the $D$ and $\bar D$ mesons
in nuclear matter in the presence of strong magnetic fields
have been recently investigated \cite{dmeson_mag}. 
These masses are calculated as arising due to their
interactions with the nucleons and the scalar mesons
in the magnetized nuclear matter.
The proton has contributions from the Landau energy
levels, in the presence of the external magnetic field. 
The effects of isospin asymmetry as well as anomalous magnetic moments
of the nucleons 
\cite{broderick1,broderick2,Wei,mao,amm,VD_SS,aguirre_fermion,aguirre_meson}
are considered for the study of the masses of $D$ mesons
in the magnetized nuclear matter.
In the present work, we study the medium modifications of the
masses of the $B$ and $\bar B$ mesons in nuclear matter in the
presence of strong magnetic fields within the chiral effective model, 
including the effects from
isospin asymmetry as well as anomalous magnetic moments
of the proton and neutron.

We organize the paper as follows. In section II, we describe the
chiral effective model used for the study of the modifications 
of the masses $B$ and $\bar B$ mesons in the strongly magnetized 
(asymmetric) nuclear matter in the present investigation.
In Section III, we discuss the results of the effects of the
magnetic field (including the effects of the anomalous
magnetic moments of the nucleons) on the $B$ and $\bar B$ masses 
in the isospin asymmetric magnetized nuclear medium. 
Section IV summarizes the findings of the present work.

\section{$B$ and $\bar B$ mesons in magnetized nuclear matter}
In the present work, we use the effective chiral model
for the study of the medium modifications of $B$ and $\bar B$
mesons in nuclear matter in the presence of magnetic field.
The Lagrangian density for the model is given as \cite{dmeson_mag}
\be
{\cal L} = {\cal L}_{kin} + \sum_{ W =X,Y,V,{\cal A},u }{\cal L}_{BW}
          + {\cal L}_{vec} + {\cal L}_0 +
{\cal L}_{scalebreak}+ {\cal L}_{SB}+{\cal L}_{mag},
\label{genlag} \ee 
which is based on a nonlinear realization of SU(3) chiral symmetry 
\cite{weinberg,coleman,bardeen} and broken scale invariance
\cite{paper3,sche1,kristof1}.
In Eq. (\ref{genlag}), 
$ {\cal L}_{kin} $ is the kinetic energy term,
$  {\cal L}_{BW}$ contains the baryon-meson interactions.
The baryon masses are generated by the baryon-scalar meson 
interaction terms in the Lagrangian density
and the parameters corresponding to these interactions are adjusted 
so as to obtain the baryon masses as their experimentally measured 
vacuum values. $ {\cal L}_{vec} $ describes the dynamical mass
generation of the vector mesons via couplings to the scalar fields
and contains additionally quartic self-interactions of the vector
fields. ${\cal L}_0 $ contains the meson-meson interaction terms
inducing the spontaneous breaking of chiral symmetry, 
${\cal L}_{scalebreak}$ is a scale invariance breaking logarithmic 
potential, $ {\cal L}_{SB} $ describes the explicit chiral symmetry
breaking. The last term, ${\cal L}_{mag}$ is the contribution 
from the magnetic field, given as
\be 
{\cal L}_{mag}=-{\bar {\psi_i}}q_i 
\gamma_\mu A^\mu \psi_i
-\frac {1}{4} \kappa_i \mu_N {\bar {\psi_i}} \sigma ^{\mu \nu}F_{\mu \nu}
\psi_i
-\frac{1}{4} F^{\mu \nu} F_{\mu \nu}.
\label{lmag}
\ee
In the above, $\psi_i$ corresponds to the $i$-th baryon.
The second term in equation (\ref{lmag}) corresponds 
to the tensorial interaction
with the electromagnetic field and is related to the
anomalous magnetic moment of the baryon (proton and neutron
for nuclear matter as considered in the present investigation). 
In this term, $\mu_N$ is the Nuclear 
Bohr magneton, given as $\mu_N=\frac{e}{2m_N}$, where 
$m_N$ is the vacuum mass of the nucleon.
$F^{\mu \nu}$ is the electromagnetic tensor given as
$F^{\mu \nu}=\partial ^\mu A^\nu-\partial ^\nu A^\mu$.
We choose the magnetic field to be uniform and along the
z-axis, i.e., $\vec B = (0,0,B)$, and take the vector 
potential to be $A^\mu =(0,0,Bx,0)$. 

For the study of $B$ and $\bar B$ mesons in the nuclear matter
in the presence of magnetic fields, we write the Lagrangian 
density in the mean field approximation (where the meson fields
are treated as classical fields), and determine these fields 
from their coupled equations of motion \cite{hartree,kristof1}. 
We use the frozen glueball approximation, i.e., neglect 
the medium dependence of the dilaton field, which is small 
compared to those of the scalar fields.
The equations of motion of the scalar fields are given as
\begin{eqnarray}
&& k_{0}\chi^{2}\sigma-4k_{1}\left( \sigma^{2}+\zeta^{2}
+\delta^{2}\right)\sigma-2k_{2}\left( \sigma^{3}+3\sigma\delta^{2}\right)
-2k_{3}\chi\sigma\zeta \nonumber\\
&-&\frac{d}{3} \chi^{4} \bigg (\frac{2\sigma}{\sigma^{2}-\delta^{2}}\bigg )
+\left( \frac{\chi}{\chi_{0}}\right) ^{2}m_{\pi}^{2}f_{\pi}
-g_{\sigma N}(\rho_{s}^{p}+ \rho_{s}^{n}) = 0,
\label{sigma}
\end{eqnarray}
\begin{eqnarray}
&& k_{0}\chi^{2}\zeta-4k_{1}\left( \sigma^{2}+\zeta^{2}+\delta^{2}\right)
\zeta-4k_{2}\zeta^{3}-k_{3}\chi\left( \sigma^{2}-\delta^{2}\right)\nonumber\\
&-&\frac{d}{3}\frac{\chi^{4}}{\zeta}+\left(\frac{\chi}{\chi_{0}} \right) 
^{2}\left[ \sqrt{2}m_{k}^{2}f_{k}-\frac{1}{\sqrt{2}} m_{\pi}^{2}f_{\pi}\right]
-g_{\zeta N}(\rho_{s}^{p}+ \rho_{s}^{n}) = 0, 
\label{zeta}
\end{eqnarray}
\begin{eqnarray}
& & k_{0}\chi^{2}\delta-4k_{1}\left( \sigma^{2}+\zeta^{2}+\delta^{2}\right)
\delta-2k_{2}\left( \delta^{3}+3\sigma^{2}\delta\right) +k_{3}\chi\delta 
\zeta \nonumber\\
& + &  \frac{2}{3} d \left( \frac{\delta}{\sigma^{2}-\delta^{2}}\right)
-g_{\delta N}(\rho_{s}^{p}- \rho_{s}^{n}) = 0, 
\label{delta}
\end{eqnarray}
where, $\rho_s^p$ and  $\rho_s^n$ are the scalar densities 
for the proton and neutron respectively. In the magnetized
nuclear matter, the expressions for these scalar densities 
are given as \cite{Wei,mao}
\begin{eqnarray}
\rho_s^p & = & \frac{eB{m_p}^*}{2\pi^2} \Bigg [ 
\sum_{\nu=0}^{\nu_{max}^{(s=1)}}
\frac {\sqrt {{m_p^*}^2+2eB\nu}+\Delta_p}{\sqrt {{m_p^*}^2+2eB\nu}}
\ln |\frac{ k_{f,\nu,1}^{(p)} + E_f^{(p)}}{\sqrt {{m_p^*}^2
+2eB\nu}+\Delta_p}|\nonumber \\
 &+&\sum_{\nu=1}^{\nu_{max}^{(s=-1)}}
\frac {\sqrt {{m_p^*}^2+2eB\nu}-\Delta_p}{\sqrt {{m_p^*}^2+2eB\nu}}
\ln |\frac{ k_{f,\nu,-1}^{(p)} + E_f^{(p)}}{\sqrt {{m_p^*}^2
+2eB\nu}-\Delta_p}|\Bigg ]
\label{scaldensityproton}
\end{eqnarray}
and
\begin{equation}
\rho_s^n =\frac{m_n^*}{4\pi^2} \sum _{s=\pm 1} 
\Bigg [ k_{f,s}^{(n)} E_f^{(n)} - 
(m_n^*+s\Delta_n)^2 \ln | \frac {k_{f,s}^{(n)}+ 
E_f^{(n)}}{m_n^*+s\Delta_n} | \Bigg].
\label{scaldensityneutron}
\end{equation}
In the expression for the scalar density for the proton as given by
equation (\ref{scaldensityproton}), one sees that there are contributions
from the Landau energy levels due to the presence of the external
magnetic field, $B$ along z-axis. $k_{f,\nu,\pm 1}^{(p)}$ 
are the Fermi momenta of protons for the Landau level, 
$\nu$ for the spin index, $s=\pm 1$
(for the spin up and down projections), 
which are related to the
Fermi energy of the proton as 
\begin{equation}
k_{f,\nu,s}^{(p)}=\sqrt { {E_f^{(p)}}^2
-\Big (
{\sqrt {{m_p^*}^2+2eB\nu}+s\Delta_p}\Big )^2}.
\end{equation}
In the scalar density for the neutron given by
equation (\ref{scaldensityneutron}), the Fermi momenta
for the neutron, for spin projection,  
$s=\pm 1$ (corresponding to the spin up (down)
projection of the neutron), are given as 
\begin{equation}
k_{f,s}^{(n)}= \sqrt { {E_f^{(n)}}^2 -
(m_n^*+s\Delta_n)^2},
\end{equation}
where $E_f^{(n)}$ is the Fermi energy of the neutron.
The effects of the anomalous magnetic moments for the
proton and neutron are encoded in the 
parameter $\Delta _{p(n)}$, which is given as
$\Delta_{p(n)} =-\frac{1}{2} \kappa_{p(n)} \mu_N B$,
where, $\kappa_i$ ($i=p,n$ for nuclear matter)
is as defined in the electromagnetic tensor term
in the Lagrangian density given by (\ref{lmag}).
The values of $\kappa_p$ and  $\kappa_n$
of $3.5856$ and $-3.8263$ are the values
of the gyromagnetic ratio arising from the 
anomalous magnetic moments of the proton and 
neutron respectively.

In the present work, we study the in-medium masses of the
open bottom mesons in the presence of a magnetic field,
for given values isospin asymmetry parameter, 
$\eta=(\rho_n-\rho_p)/(2\rho_B)$, where $\rho_B$ 
is the baryon density, and the number densities for
the proton and the neutron are given as
\begin{equation}
\rho_p=\frac{eB}{4\pi^2} \Bigg [ 
\sum_{\nu=0}^{\nu_{max}^{(s=1)}} k_{f,\nu,1}^{(p)} 
+\sum_{\nu=1}^{\nu_{(max)}^{(s=-1)}} k_{f,\nu,-1}^{(p)} 
\Bigg]
\end{equation}
and 
\begin{equation}
\rho_{n}= \frac{1}{4\pi^2} \sum _{s=\pm 1}
\Bigg \{ \frac{2}{3} {k_{f,s}^{(n)}}^3
+s\Delta_n \Bigg[ (m_n^*+s\Delta_n) k_{f,s}^{(n)}
+{E_f^{(n)}}^2 \Bigg( arcsin \Big (
\frac{m_n^*+s\Delta_n}{E_f^{(n)}}\Big)-\frac{\pi}{2}\Bigg)\Bigg]
\Bigg \}.
\end{equation}
respectively.
%
As has already been mentioned, in the present work,
the in-medium masses of the $B$ and $\bar B$ in the magnetized 
nuclear matter are studied in a chiral effective model,
which is a generalization of a chiral SU(3) model,
to the charm and bottom sectors, so as to derive the
interactions of the charm mesons
\cite{amdmeson,amarindamprc,amarvdmesonTprc,amarvepja,DP_AM_Ds},
and bottom mesons \cite{DP_AM_bbar,DP_AM_Bs}
with the light hadronic sector. 
The interaction Lagrangian modifying the $B$ and $\bar B$ mesons 
can be written as \cite{DP_AM_bbar}
\begin{eqnarray}
\cal L _{BN} & = & -\frac {i}{8 f_B^2} \Big [3\Big (\bar p \gamma^\mu p
+\bar n \gamma ^\mu n \Big) 
\Big \{\Big( (\partial_\mu {B^+})  B^- -
B^+ (\partial_\mu B^-) \Big )
+\Big( (\partial_\mu {{B^0}}) {\bar B}^0
- {B^0} (\partial_\mu \bar B^0)  \Big ) \Big\}
\nonumber \\
& +&
\Big (\bar p \gamma^\mu p -\bar n \gamma ^\mu n \Big) 
\Big \{ \Big( (\partial_\mu {B^+})  B^- -
B^+ (\partial_\mu B^-) \Big )
-\Big( (\partial_\mu {{B^0}}) {\bar B}^0
- {B^0} (\partial_\mu \bar B^0)  \Big )\Big\}
\Big ]
\nonumber \\
 &+ & \frac{m_B^2}{2f_B} \Big [ 
(\sigma +\sqrt 2 \zeta_b)\big ((B^+ B^-) +
{ B^0} \bar B^0 \big )
 +\delta \big ( (B^+ B^-) -{ B^0} \bar B^0 )
\big ) \Big ] \nonumber \\
& - & \frac {1}{f_B}\Big [ 
(\sigma +\sqrt 2 \zeta_b )
\Big ( (\partial _\mu {B^+})(\partial ^\mu {B^-}) + 
(\partial_\mu {{B^0}})(\partial ^\mu  {{\bar B}^0})
\Big)
\nonumber \\
 & + & \delta
\Big ( (\partial _\mu {B^+})(\partial ^\mu {B^-}) - 
(\partial_\mu {{B^0}})(\partial ^\mu  {{\bar B}^0})
\Big)
\Big ]
\nonumber \\
&+ & \frac {d_1}{2 f_B^2}(\bar p p +\bar n n 
 )\big ( (\partial _\mu {B^+})(\partial ^\mu {B^-})
+(\partial _\mu {B^0})(\partial ^\mu {\bar {B^0}})
\big )
\nonumber \\
&+& \frac {d_2}{4 f_B^2} \Big [
3(\bar p p+\bar n n))\big ( 
(\partial _\mu {B^+})(\partial ^\mu {B^-})
+(\partial _\mu {B^0})(\partial ^\mu {\bar {B^0}})
\big )
\nonumber \\
 &+&  (\bar p p -\bar n n) \big ( 
(\partial _\mu {B^+})(\partial ^\mu {B^-})
-(\partial _\mu {B^0})(\partial ^\mu {\bar {B^0}})
\big )
\Big ]
\label{lagbn}
\end{eqnarray}
In (\ref{lagbn}), the first term is the vectorial Weinberg Tomozawa
interaction term, which is attractive for $\bar B$ mesons,
but repulsive for the $B$ mesons. 
The second term is the scalar meson exchange
term, which is attractive for both $B$ and $\bar B$ mesons.
The third, fourth and fifth terms comprise
the range term in the chiral model. 
The parameters $d_1$ and $d_2$ in the last two terms of the
interaction Lagrangian given by (\ref{lagbn}) are determined
by fitting to the empirical values of the KN scattering lengths
\cite{thorsson,juergen,barnes}
for I=0 and I=1 channels \cite{isoamss1,isoamss2}.

The dispersion relations for the $B$ and $\bar B$ mesons 
are obtained from the Fourier transformations of the
equations of motion of these mesons. These are given as
\begin{equation}
-\omega^2+ {\vec k}^2 + m_{B(\bar B)}^2
 -\Pi_{B(\bar B)}(\omega, |\vec k|)=0,
\label{dispbbar}
\end{equation}
where $\Pi_{B(\bar B)}$ denotes the self energy 
of the $B$ ($\bar B$) meson in the medium.
For the $B$ meson doublet ($B^+$,$B^0$), and $\bar B$ meson
doublet ( $B^-$, ${\bar B}^0$), the self energies are given by
\begin{eqnarray}
\Pi _{B} (\omega, |\vec k|) &= & -\frac {1}{4 f_B^2}\Big [3 (\rho_p +\rho_n)
\pm (\rho_p -\rho_n) \big)
\Big ] \omega \nonumber \\
&+&\frac {m_B^2}{2 f_B} (\sigma ' +\sqrt 2 {\zeta_b} ' \pm \delta ')
\nonumber \\ & +& \Big [- \frac {1}{f_B}
(\sigma ' +\sqrt 2 {\zeta_b} ' \pm \delta ')
+\frac {d_1}{2 f_B ^2} (\rho_s ^p +\rho_s ^n)\nonumber \\
&+&\frac {d_2}{4 f_B ^2} \Big (3({\rho^s} _p +{\rho^s} _n)
\pm   ({\rho^s} _p -{\rho^s} _n) \Big ) \Big ]
(\omega ^2 - {\vec k}^2),
\label{selfb}
\end{eqnarray}
and
\begin{eqnarray}
\Pi _{\bar B} (\omega, |\vec k|) 
&= & \frac {1}{4 f_B^2}\Big [3 (\rho_p +\rho_n)
\pm (\rho_p -\rho_n) \Big ] \omega\nonumber \\
&+&\frac {m_B^2}{2 f_B} (\sigma ' +\sqrt 2 {\zeta_b} ' \pm \delta ')
\nonumber \\ & +& \Big [- \frac {1}{f_B}
(\sigma ' +\sqrt 2 {\zeta_b} ' \pm \delta ')
+\frac {d_1}{2 f_B ^2} (\rho_s ^p +\rho_s ^n
)\nonumber \\
&+&\frac {d_2}{4 f_B ^2} \Big (3({\rho^s} _p +{\rho^s} _n)
\pm   ({\rho^s} _p -{\rho^s} _n) \Big ]
(\omega ^2 - {\vec k}^2),
\label{selfbbar}
\end{eqnarray}
where the $\pm$ signs refer to the $B^+$ and $B^0$ respectively
in equation (\ref{selfb}) and to the $B^-$  and $\bar {B^0}$ mesons
respectively in equation (\ref{selfbbar}).
In equations (\ref{selfb}) and (\ref{selfbbar}), $\sigma'(=\sigma-\sigma _0)$,
 ${\zeta_b}'(=\zeta_b-{\zeta_b}_0)$ and  $\delta'(=\delta-\delta_0)$
are the fluctuations of $\sigma$, $\zeta_b$, and $\delta$,
from their vacuum expectation values. In the present calculations,
we neglect the fluctuation
of $\zeta_b \sim \langle \bar b b \rangle$, due to the
reason that the fluctuation of the heavy quark
condensates are small in the medium \cite{DP_AM_bbar}.

The masses of the charged open bottom mesons, $B^{\pm}$ have 
an additional positive mass shift due to the presence of the
magnetic field, which retaining only the lowest Landau level,
are given as
\begin{equation}
m^{eff}_{B^\pm}=\sqrt {{m^*_{B^\pm}}^2 +|eB|},
\label{mbpm_landau}
\end{equation}
whereas for the neutral $B(\bar B)$ mesons, namely,
$B^0 (\bar {B^0})$, the effective masses are given as
$m^{eff}_{B^0 (\bar {B^0})}=m^*_{B^0 (\bar {B^0})}$,
which are the solutions for $\omega$ at $|\vec k|=0$,
of the dispersion relations given by equation 
(\ref{dispbbar}), using the self-energies for $B^0$ and
$\bar {B^0}$ as given by (\ref{selfb}) and (\ref{selfbbar}).

In the next section, we shall discuss the results for 
the $B(\bar B)$-meson mass modifications in symmetric
as well as asymmetric nuclear
matter in the presence of an external magnetic field, $B$, as
obtained in the present chiral effective model.
The in-medium masses of the $B$ and $\bar B$ mesons are
studied accounting for the effects of the anomalous magnetic
moments of the nucleons.

\begin{figure}
\includegraphics[width=16cm,height=16cm]{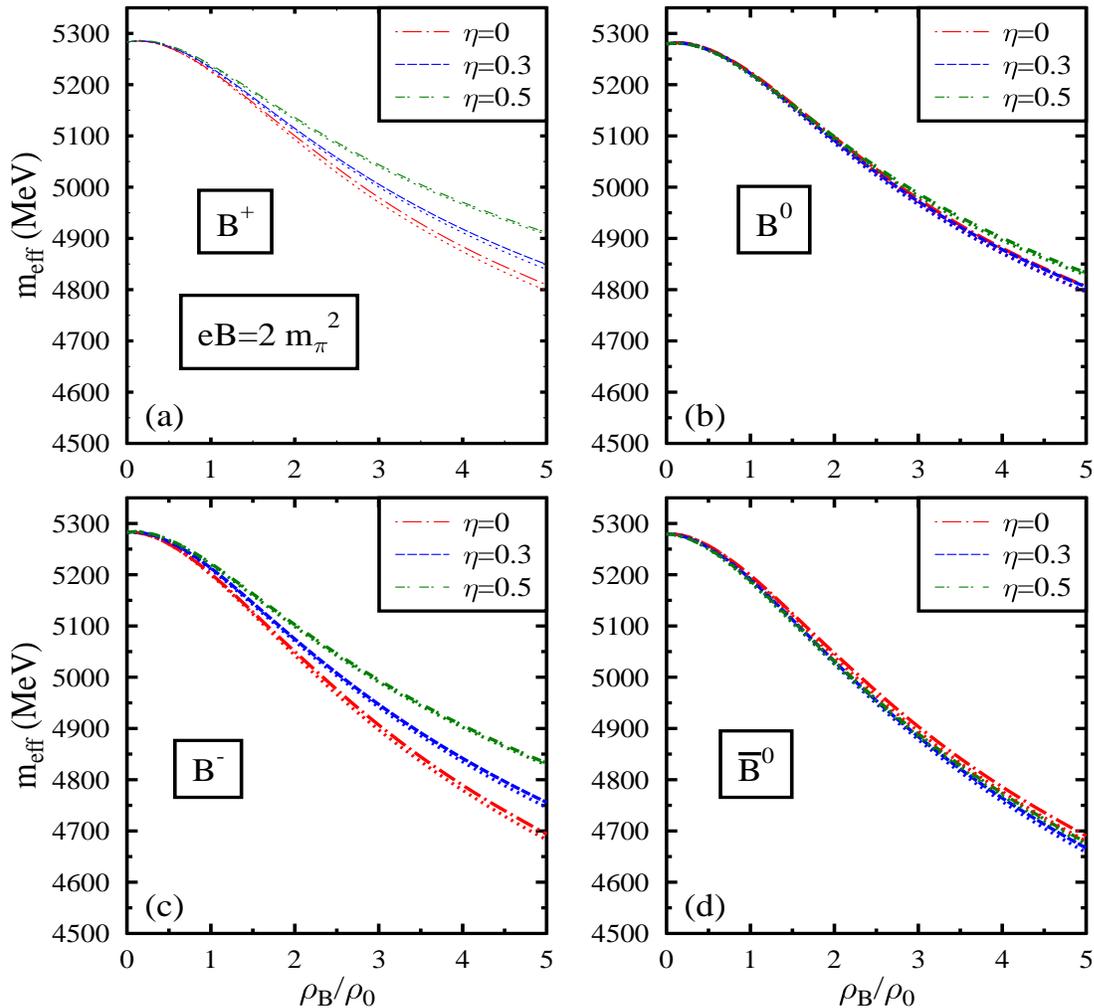}
\caption{
The in-medium masses of the $B$ and $\bar B$ mesons plotted
as functions of the baryon density (in units of 
nuclear matter saturation density) for various 
values of isospin asymmetry parameter, $\eta$,
for value of  the magnetic field, $eB=2 m_\pi^2$,
accounting for the effects of anomalous magnetic moment.
These results are compared to the case when the
effects of the anomalous magnetic moment
are not taken into account (shown as dotted line).
\label{mbbar_2mpi2}
}
\end{figure}

\begin{figure}
\includegraphics[width=16cm,height=16cm]{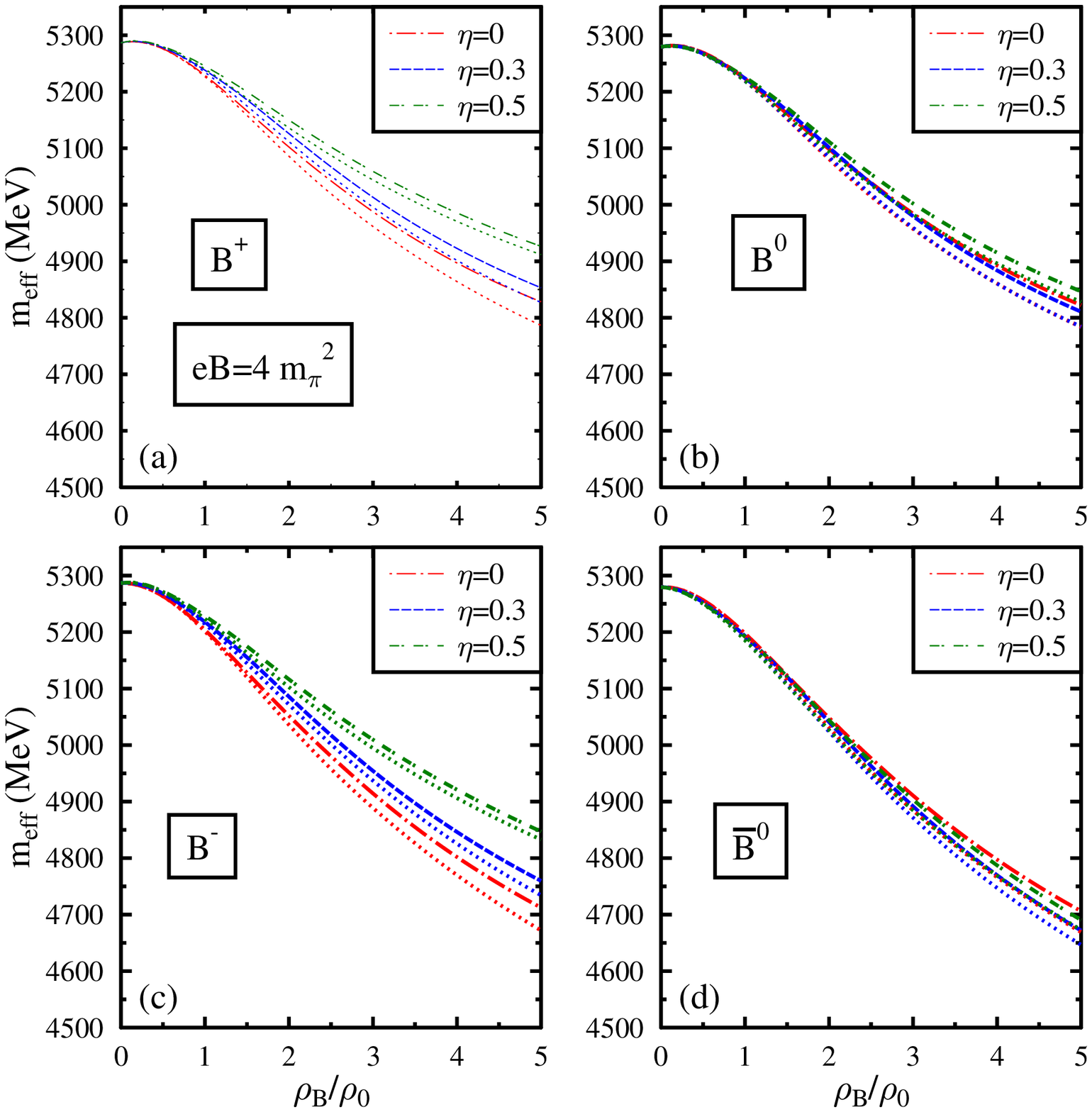}
\caption{
The in-medium masses of the $B$ and $\bar B$ mesons plotted
as functions of the baryon density (in units of 
nuclear matter saturation density) for various 
values of isospin asymmetry parameter, $\eta$,
for value of  the magnetic field, $eB=4 m_\pi^2$,
accounting for the effects of anomalous magnetic moment.
These results are compared to the case when the
effects of the anomalous magnetic moment
are not taken into account (shown as dotted line).
\label{mbbar_4mpi2}
}
\end{figure}
 
\begin{figure}
\includegraphics[width=16cm,height=16cm]{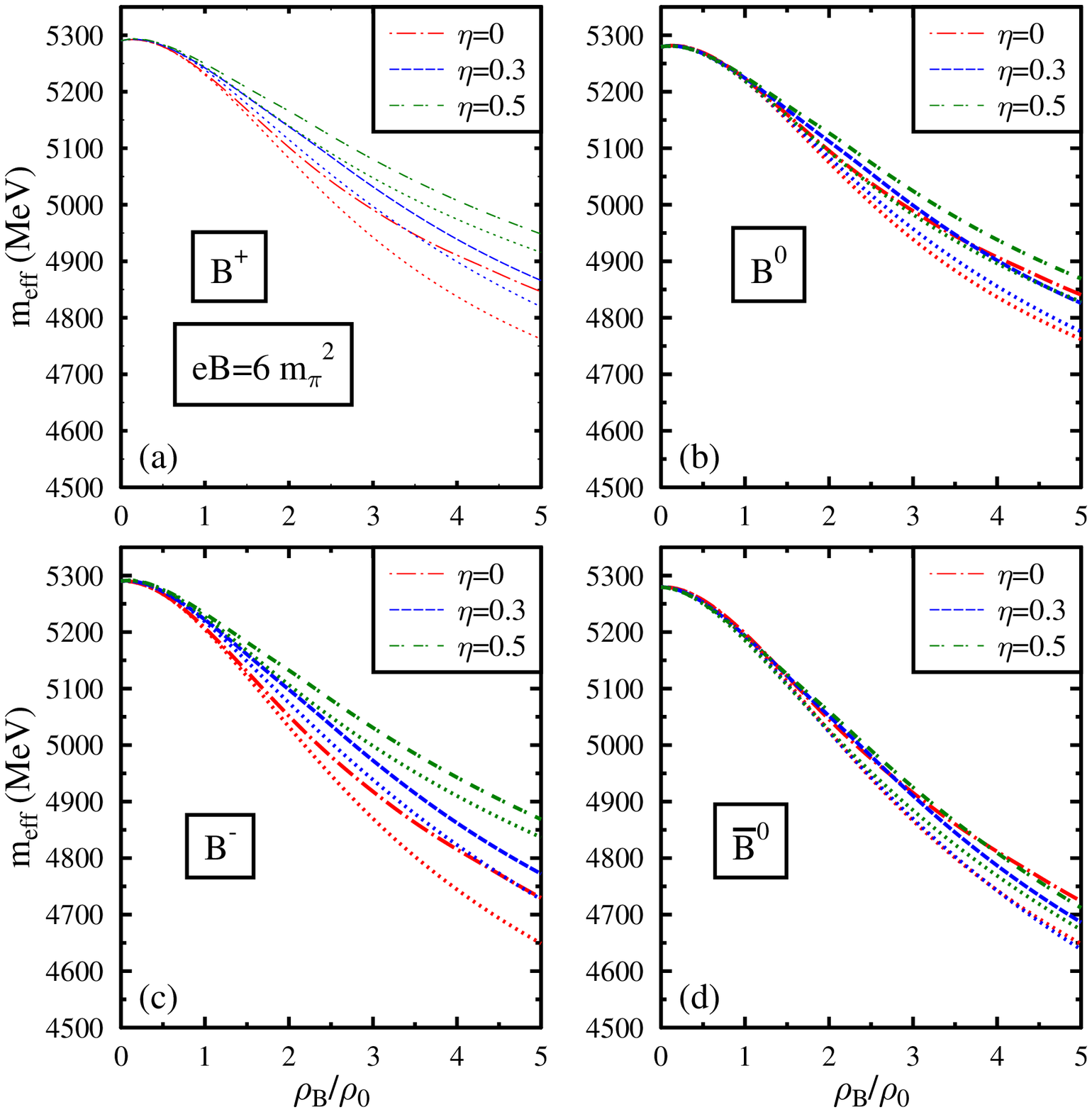}
\caption{
The in-medium masses of the $B$ and $\bar B$ mesons plotted
as functions of the baryon density (in units of 
nuclear matter saturation density) for various 
values of isospin asymmetry parameter, $\eta$,
for value of  the magnetic field, $eB=6 m_\pi^2$,
accounting for the effects of anomalous magnetic moment.
These results are compared to the case when the
effects of the anomalous magnetic moment
are not taken into account (shown as dotted line).
\label{mbbar_6mpi2}
}
\end{figure}

\begin{figure}
\includegraphics[width=16cm,height=16cm]{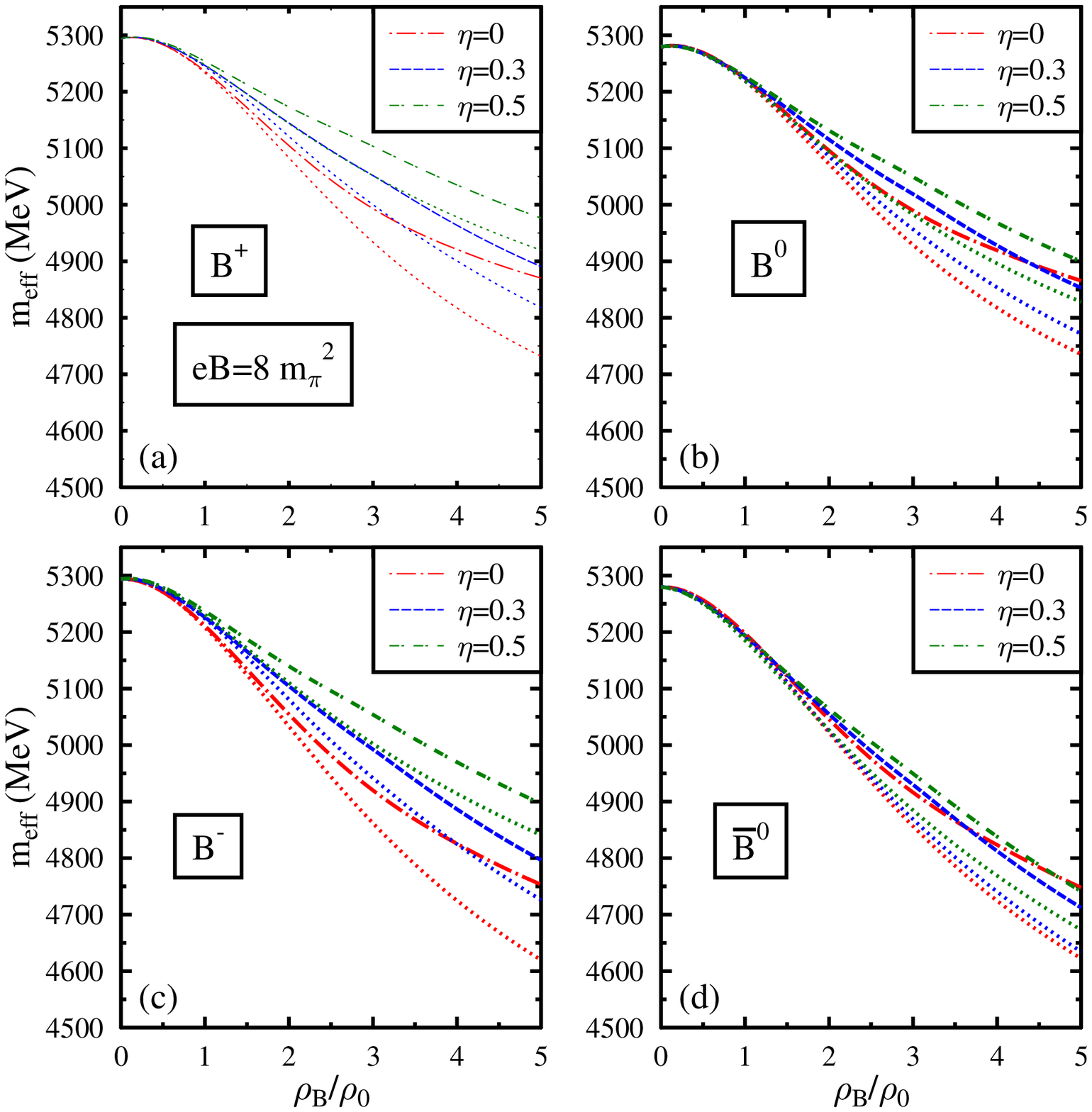}
\caption{
The in-medium masses of the $B$ and $\bar B$ mesons plotted
as functions of the baryon density (in units of 
nuclear matter saturation density) for various 
values of isospin asymmetry parameter, $\eta$,
for value of  the magnetic field, $eB=8 m_\pi^2$,
accounting for the effects of anomalous magnetic moment.
These results are compared to the case when the
effects of the anomalous magnetic moment
are not taken into account (shown as dotted line).
\label{mbbar_8mpi2}
}
\end{figure}

\begin{figure}
\includegraphics[width=16cm,height=16cm]{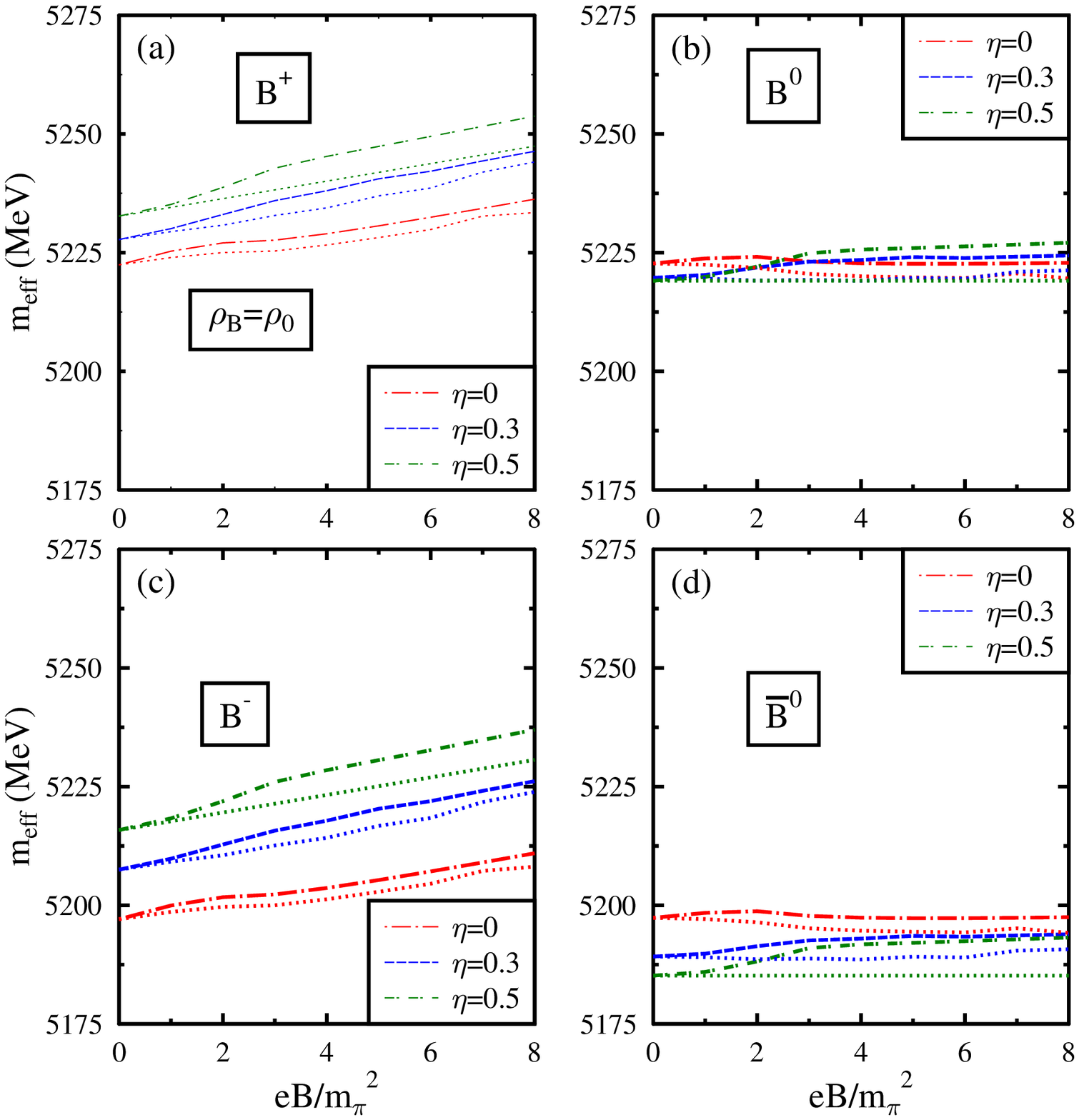}
\caption{
The effective masses of $B$ ($B^+$,$B^0$) and $\bar B$ 
($B^-$, $\bar {B^0}$) mesons 
in MeV plotted as functions of 
$eB/{m_\pi^2}$, 
for baryon density, $\rho_B=\rho_0$,
with different values of  isospin asymmetry parameter, $\eta$,
accounting for the effects of
the anomalous magnetic moments (AMM) for the nucleons.
The results are compared with the case of not accounting for the
anomalous magnetic moments (shown as dotted lines).
\label{mbbbar_mag_rhb0}
}
\end{figure}

\begin{figure}
\includegraphics[width=16cm,height=16cm]{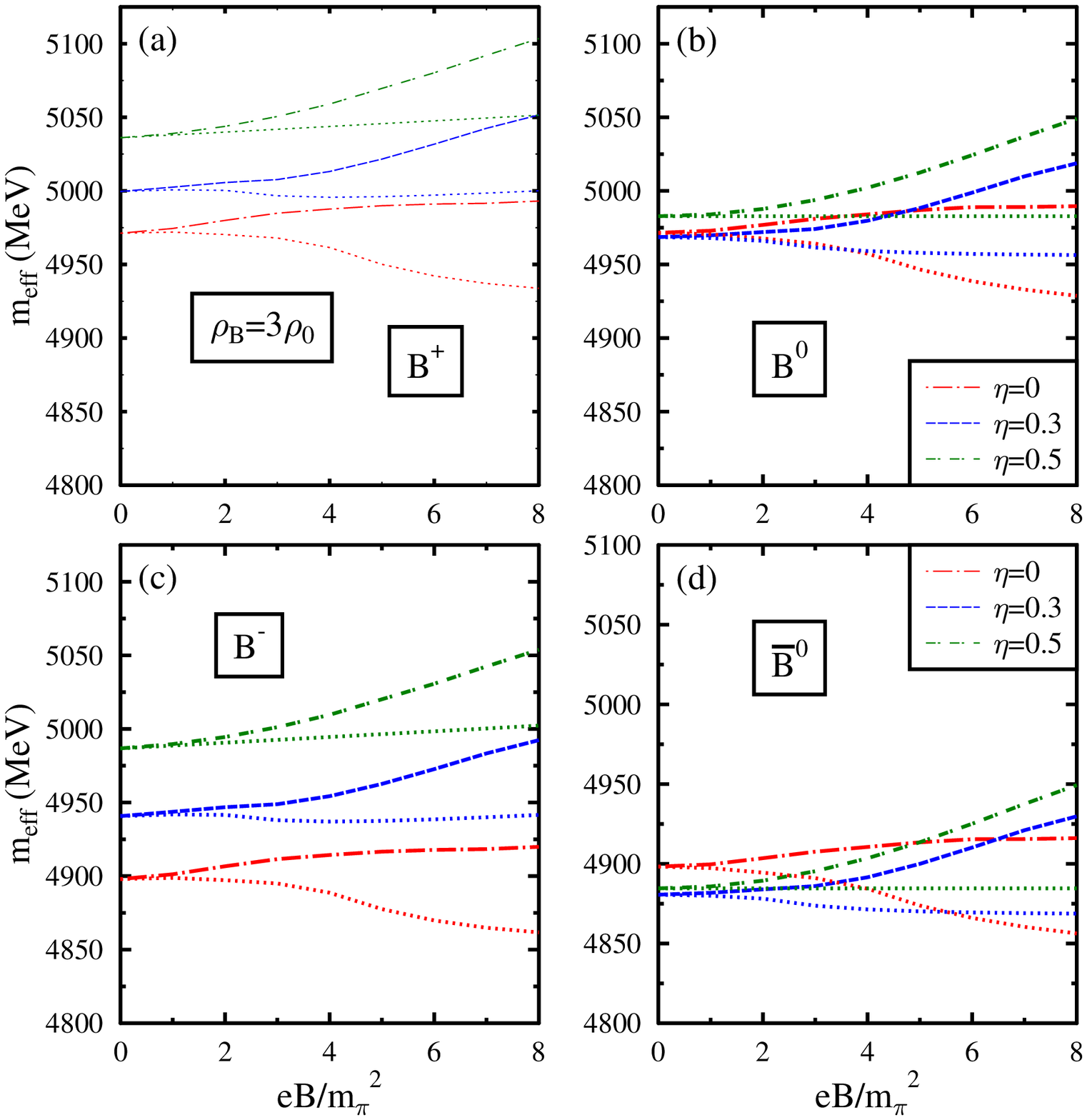}
\caption{
The effective masses of $B$ ($B^+$,$B^0$) and $\bar B$ 
($B^-$, $\bar {B^0}$) mesons 
in MeV plotted as functions of 
$eB/{m_\pi^2}$, 
for baryon density, $\rho_B=3\rho_0$,
with different values of  isospin asymmetry parameter, $\eta$,
accounting for the effects of
the anomalous magnetic moments (AMM) for the nucleons.
The results are compared with the case of not accounting for the
anomalous magnetic moments (shown as dotted lines).
\label{mbbbar_mag_3rhb0}
}
\end{figure}

\begin{figure}
\includegraphics[width=16cm,height=16cm]{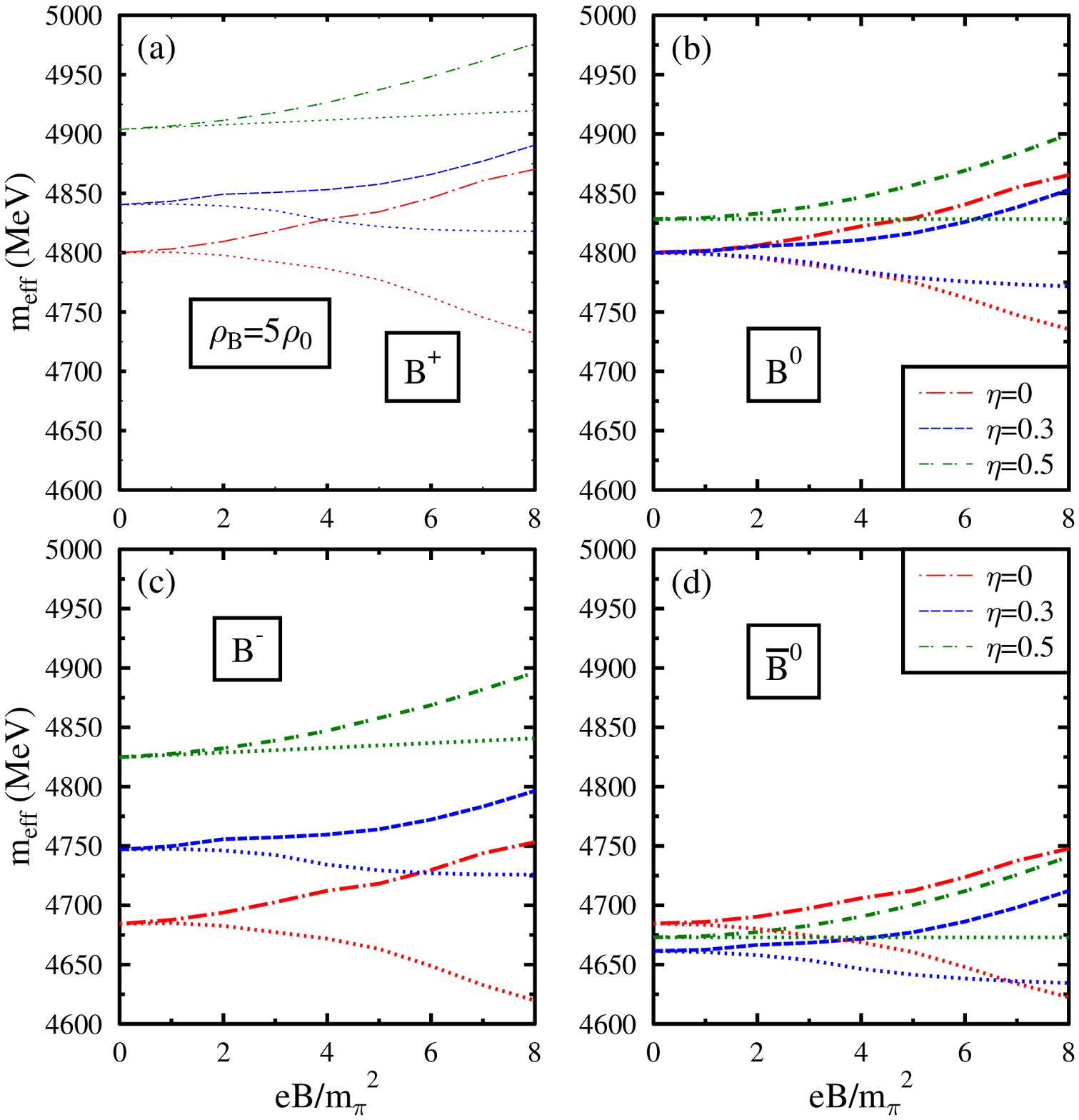}
\caption{
The effective masses of $B$ ($B^+$,$B^0$) and $\bar B$ 
($B^-$, $\bar {B^0}$) mesons 
in MeV plotted as functions of 
$eB/{m_\pi^2}$, 
for baryon density, $\rho_B=5\rho_0$,
with different values of  isospin asymmetry parameter, $\eta$,
accounting for the effects of
the anomalous magnetic moments (AMM) for the nucleons.
The results are compared with the case of not accounting for the
anomalous magnetic moments (shown as dotted lines).
\label{mbbbar_mag_5rhb0}
}
\end{figure}

\section{Results and Discussions}
\label{results}

We study the medium modifications of the masses of open bottom mesons 
in nuclear matter
in presence of an external magnetic field using a chiral effective model.
The modifications of the masses of the $B$ and $\bar B$ mesons
arise from their interactions with the protons 
and neutrons, as well as the scalar mesons (isoscalar non-strange
$\sigma$ meson, isoscalar strange $\zeta$ meson and
isovector $\delta$ meson). The calculations are carried out
within the frozen glueball approximation, where the medium
modifications of the scalar dilaton
field, $\chi$, which mimicks the gluon condensates of QCD,
are not taken into account.
We study the mass modifications in $B$ and $\bar B$ mesons 
accounting for the anomalous magnetic moments (AMM) 
of the nucleons and compared to the case when 
AMM effects are not taken into consideration. 
The presence of magnetic field gives rise to Landau energy 
levels for charged baryon in the medium i.e. proton.
The in-medium masses of the $B$ mesons are calculated 
from the dispersion relation (\ref{dispbbar}),
with their self-energies being given by (\ref{selfb}).
As already stated, the charged $B$ mesons undergo 
further mass modification in presence of magnetic field 
which is given by (\ref{mbpm_landau}), retaining contribution
due to lowest Landau energy level. 

In the figures \ref{mbbar_2mpi2}, \ref{mbbar_4mpi2}, \ref{mbbar_6mpi2},
\ref{mbbar_8mpi2}, 
we show the masses of the $B$ and $\bar B$ mesons,
plotted as functions of the baryon  density in units of nuclear 
saturation density, $\rho_0$, for the values of the
$eB$ as $2m_\pi^2$, $4m_\pi^2$, $6m_\pi^2$, $8m_\pi^2$, 
respectively.
As can be seen from the expression for the self energy
of the $B$ mesons given by (\ref{selfb}),
the isospin symmetric part of the Weinberg-Tomozawa term,
is repulsive for the $B$ mesons (similar to the case
of the $K$ ($\bar D$) mesons), whereas this term is
attractive for the $\bar B$ mesons (similar
to the $\bar K$ ($D$) mesons) in nuclear matter
\cite{isoamss,isoamss1,isoamss2,amdmeson,amarindamprc,
amarvdmesonTprc,amarvepja}.
The isospin symmetric part of Weinberg-Tomozawa term thus leads to
an increase in the masses of both $B^+$ and $B^0$ mesons,
whereas, the isospin asymmetric 
part of the Weinberg-Tomozawa term leads to a positive 
(negative) shift in the mass of the $B^0$ ($B^+$) meson,
in the isospin asymmetric nuclear medium. 
The scalar exchange term in the isospin symmetric nuclear matter
is attractive for both the $B$ and $\bar B$ mesons.
The last three terms of the self energies for the 
$B$ and $\bar B$ mesons, correspond to the range terms,
which are repulsive for the first range term and attractive
for the other two terms (the $d_1$ and $d_2$ terms)
for isospin symmetric nuclear matter. Accounting for 
the contributions from all the terms (Weinberg-Tomozawa,
scalar exchange and the range terms),  as well as for 
a mass shift for the charged $B^\pm$ mesons due to
Landau quantization, there is observed to be drop
in the masses of the $B$ and $\bar B$ mesons
with increase in density.
There is observed to be increase in the 
in-medium masses of the $B$ as well as $\bar B$ mesons,
when the effects of the anomalous magnetic moments
are taken into account, as compared to when these
are not taken into consideration.
These effects are observed to be larger for higher values of the
magnetic field.
 
For the $\bar B$ mesons, the isospin symmetric part of the Weinberg-Tomozawa
term is attractive and leads to drop in their masses. With inclusion
of isospin asymmetry in the medium, the 
mass of $\bar {B^0}$ is further decreased while that of $B^-$
has a positive contribution from isospin asymmetry.
The behaviour of the $B^-$ meson mass is very similar
to the mass of $B^+$ with isospin asymmetry, but the 
drop in the mass of $B^-$ is much larger
than that of $B^+$ meson, due to the attractive
Weinberg-Tomozawa contribution.

We next discuss, specifically, the effects of the magnetic field 
on the masses of the $B$ and $\bar B$ mesons 
for specific values of density and isospin asymmetry of 
the nuclear matter. These masses are plotted
for densities $\rho_0$, $3\rho_0$ and $5\rho_0$ in figures  
\ref {mbbbar_mag_rhb0}, \ref {mbbbar_mag_3rhb0} and
\ref {mbbbar_mag_5rhb0} respectively. 
For baryon density, $\rho_0$, as shown in figure \ref {mbbbar_mag_rhb0}, 
for symmetric nuclear matter, 
while accounting for the effects of anomalous magnetic moments (AMM)
of the nucleons, the masses of $B^0$ as well as $\bar {B^0}$ 
are observed to show an initial increase with increase in magnetic field
(upto $eB \sim 2-3 m_\pi^2$),
followed by negligible change wih further increase in
the magnetic field. In the absence of the AMM effects,
in symmetric nuclear matter, the 
masses of these neutral $B$ and $\bar B$ mesons show a drop
at small values of the magnetic field and marginal modifications
as the magnetic field is further increased. However, the 
magnetic field is observed to have substantial influence on the masses
of charged mesons $B^+$ and $B^-$ mesons (of around 15 MeV)
in symmetric nuclear matter at $\rho_B=\rho_0$, 
due to the Landau quantization effects
in the presence of magnetic field. 
The effects of the AMM are observed to lead to higher masses
of the $B$ as well as $\bar B$ mesons as compared to when these
effects are not taken into account. 
For asymmetric nuclear matter, with AMM, the masses of $B^0$ and $\bar {B^0}$
undergo a slight increase ($\sim$ $5-8$ MeV) as magnetic field increases 
from zero to 8$m_\pi^2$, for $\rho_B=\rho_0$. The masses of charged mesons, 
$B^+$ and $B^-$ are observed to increase appreciably
(by about $15-20$ MeV at $\rho_B=\rho_0$), 
for a similar increase in the magnetic field,
due to the Landau quantization effects. 
There is no modification of the
neutral $B^0$ and $\bar {B^0}$ meson masses for $\eta$=0.5
(when there are only neutrons in the system), due to the magnetic field,
when the AMM effects are not taken into account, as the magnetic field 
effects for the neutrons are only due to the anomalous magnetic moment
of the neutron.
At higher densities of $\rho_B$ as 3$\rho_0$ and
5$\rho_0$, shown in figures \ref {mbbbar_mag_3rhb0}
and \ref {mbbbar_mag_5rhb0},  
accounting for the effects of AMM is observed to increase
the masses of $B$ and $\bar B$ mesons and these are observed
to be larger at higher isospin asymmetry
in the nuclear matter. 
In asymmetric nuclear matter whenever isospin asymmetry parameter,
$\eta$ is different from 0.5 and if we neglect the effects of AMM, 
then the masses of these mesons are observed to drop as we increase
the magnetic field. With AMM effects, at higher densities
these masses are observed to rise with magnetic field,
and, this increase is seen to be more appreciable for 
asymmetric nuclear matter.
The difference between the masses, with and without considering the
effects of AMM, is observed to be larger for higher values of magnetic
field.

The $D$ and $\bar D$ mesons have been investigated in 
asymmetric nuclear matter in the presence of strong magnetic fields
\cite{dmeson_mag}
within the chiral effective model used in the present investigation.
We compare the effects of the magnetic field  
on the masses of the open bottom mesons studied in the present work
with the effects on the masses of the $D$ and $\bar D$ mesons.
The $\bar B$ ($D$) mesons have larger drop 
in their masses in the nuclear medium as compared to the $B$ ($\bar D$),
arising mainly due to the Weinberg-Tomozawa interaction
which is attractive for the former and repulsive for the latter.
In the absence of magnetic field, in the asymmetric nuclear matter,
the masses of the $B^+$ and $B^0$ mesons behave like
the  $\bar {D^0}$ and $D^-$ masses respectively, and, the 
masses of the $B^-$ and $\bar {B^0}$ mesons behave like
$D^0$ and $D^+$ masses.
In the presence of magnetic fields, there is additional (positive)
mass shifts for the charged $D^\pm$ as well as $B^\pm$ mesons.
However, due to the much larger mass of the $B$ and $\bar B$ mesons
as compared to the $D$ and $\bar D$ mesons, the mass shift 
is much more appreciable for the charged open charm mesons
as compared to the charged open bottom mesons.
At low densities ($\rho_B \sim \rho_0$), the changes in 
the masses of neutral $B$ and $D$ mesons are marginal
as magnetic field is increased (with modifications
upto around 5 MeV for the $D^0$ and $\bar {D^0}$
mesons and upto around 8 MeV for  the $B^0$ 
and $\bar {B^0}$ mesons). At higher densities,
accounting the effects of anomalous magnetic moment, 
the masses of all open bottom and open charm
mesons show steady increase as magnetic field increases. 
In the absence of AMM effects, the masses of the $D$ and
$\bar D$, as well as $B$ and $\bar B$ mesons 
are observed to be smaller than the case when
the AMM effects are taken into account and
the difference in the masses in the two situations
is observed to be larger at higer densities.


The effects of the isospin asymmetry as well as magnetic 
fields on the masses of the $B$ and $\bar B$ mesons
are observed to be large at high densities.
The effects of the anomalous magnetic moments (AMM) are observed
to be large at high values of the magnetic
fields leading to an increase in the masses of the 
$B$ and $\bar B$ mesons as compared to when these effects
are not taken into account. The AMM effects are observed
to be more prominent in the isospin symmetric system, and 
becomes smaller (but still appreciable) with increase
in isospin asymmetry in the nuclear medium.

\section{summary}
To summarize, we have studied the medium  modifications of the masses
 of the open bottom mesons in isospin asymmetric nuclear matter 
in presence of strong magnetic fields with emphasis on the 
behaviors when AMM of the nucleons are taken into consideration.
We have used a chiral effective model which is generalized from SU(3) 
to SU(5) in order to describe the interactions of the open bottom mesons 
with the hadrons in the medium. Due to their interactions with the 
nucleons and the scalar fields, their masses are observed to be modified. 
In presence of external magnetic fields the number densities and 
scalar densities of protons have contributions from the Landau 
energy levels. The magnetic fields also show effects through 
the anomalous magnetic moments of the nucleons which we have studied
in the present work. The effects of the isospin asymmetry
as well as magnetic fields are observed to be large
for both the $B$ and $\bar B$ meson doublets,
especially at high densities. The isospin asymmetric effects
being large at high densities should have
observable effects in the ratios of  $B^+/B^0$ and
$B^-/{\bar {B^0}}$, as well as in the decay widths of
bottomonium states to $B\bar B$, in asymmetric heavy 
ion collisions planned at Compressed baryonic matter (CBM)
experiments at FAIR at the future facility at GSI.  
 
\acknowledgements
One of the authors (AM) is grateful to the Institut 
f\"ur Theoretische Physik, Universitaet Frankfurt,
for warm hospitality and acknowledges financial 
support from Alexander von Humboldt Stiftung 
when this work was initiated. 
Amal Jahan CS acknowledges the support
towards this work from Department of Science and Technology, 
Govt of India, via  INSPIRE  fellowship scheme 
(Ref. No. DST/INSPIRE/03/2016/003555).

\end{document}